\documentclass[twocolumn,showpacs,preprintnumbers,prb,amsmath,amssymb,floatfix]{revtex4} 
\usepackage{graphicx}
\usepackage{dcolumn}
\usepackage{bm}

\begin{document} 

\setlength{\topmargin}{0in}

\title{A Neutron Elastic Diffuse Scattering Study of PMN} 
\author{Guangyong Xu}
\author{G.~Shirane} 
\affiliation{Physics Department, Brookhaven National Laboratory, Upton, 
New York 11973}
\author{J.~R.~D.~Copley}
\author{P.~M.~Gehring}
\affiliation{NIST Center for Neutron Research, National Institute of Standards
and Technology, Gaithersburg, Maryland 20899}
\date{\today} 
 
\begin{abstract} 
We have performed elastic diffuse neutron scattering studies on the relaxor 
Pb(Mg$_{1/3}$Nb$_{2/3}$)O$_3$ (PMN). The measured intensity
distribution near a (100) Bragg peak in the (hk0) scattering plane assumes
the shape of a butterfly with extended intensity in the (110) and (1$\bar{1}$0)
directions. The temperature dependence of the 
diffuse scattering shows that both the size of the polar nanoregions (PNR) 
and the integrated diffuse intensity increase with cooling even for temperatures
below the Curie temperature $T_C \sim 213$~K. 

\end{abstract} 
 

\maketitle 

\section{Introduction}
 
Pb(Mg$_{1/3}$Nb$_{2/3}$)O$_3$ (PMN) is a typical relaxor ferroelectric that has
a broad and strongly frequency-dependent dielectric constant. 
This system has attracted much attention in the last few years, 
because of the discovery of an ultrahigh 
piezoelectric response in  solid solutions with PbTiO$_3$ (PT) near the 
morphotropic phase boundary (MPB)~\cite{PZT1}.
Pure PMN is considered to be a prototype relaxor, and has been 
studied extensively in recent years~\cite{Ye_Review}. 
Nevertheless, certain aspects of this system are not fully understood. 
Unlike one of its close analogues, Pb(Zn$_{1/3}$Nb$_{2/3}$)O$_3$ (PZN), 
PMN remains cubic below the (field-induced) Curie 
temperature $T_C \approx 213$~K~\cite{Bonneau,Husson}, and was believed to 
exhibit no macroscopic ferroelectric phase transition in zero field.  

Efforts to understand the dynamical properties of relaxors have focused on 
the polar nanoregions (PNR) present in these compounds.  
Burns and Dacol~\cite{Burns} found that on cooling such systems, 
PNR start to form at temperatures a few hundred degrees above 
$T_C$. This temperature was later called the ``Burns Temperature'' $T_d$, 
$T_d\approx 620$~K for PMN. 
Neutron scattering measurements~\cite{PMN_neutron,PMN_neutron2} 
have demonstrated that diffuse scattering appears in PMN between 600~K and 
650~K, consistent with previous optical measurements by Burns and 
Dacol. The diffuse intensity that develops in relaxors below $T_d$ 
has been identified with the PNR. Various neutron and x-ray measurements
on diffuse scattering have been carried out in order to investigate how PNR are 
formed, and to determine average sizes and polarizations (atomic shifts) 
at different 
temperatures~\cite{PMN_diffuse,PMN_diffuse2,PMN_xraydiffuse,PZN_diffuse}. 
Phonon contributions have always been a potential source of 
contamination in diffuse scattering measurements. In this paper, we present,
and attempt to interpret, elastic diffuse scattering measurements 
on pure PMN.

\section{Experiment}

The experiment was carried out using the Disk Chopper
Spectrometer (DCS)~\cite{DCS} at the NIST Center for Neutron
Research. The sample was a high quality
single crystal of PMN with $\sim 0.1^\circ$ mosaic, and a mass of 3.25~g,
grown at the Simon Fraser University in Canada. 
A number of neutron scattering studies using this
crystal have already been published~\cite{PMN_softmode,PMN_diffuse,Waki1,Waki2}.
The room temperature lattice parameter is $a=4.04$~\AA. 
At each of several temperatures, 
time-of-flight spectra were collected for each of at least 51
successive crystal orientations $0.5^\circ$ apart using 325 detectors
with active dimensions in and normal to the scattering plane of $\sim$ 31~mm
and 400~mm respectively. The DCS detectors are located 4000~mm from the
sample at scattering angles from $5^\circ$ to $140^\circ$. At an 
incident neutron wavelength of 5.5~\AA, the FWHM (full width at half
maximum) of the elastic resolution function was $\sim$ 0.085meV, 
and the time between pulses at the sample is 9~ms,
effectively eliminating frame overlap. Data were collected near a (100)
Bragg peak in the (hk0) scattering plane. The choice of a low $Q$ reflection 
helps minimize phonon and multiphonon contributions.

\section{Results and Discussion}

\begin{figure}[ht]
\includegraphics[angle=90,width=\linewidth]{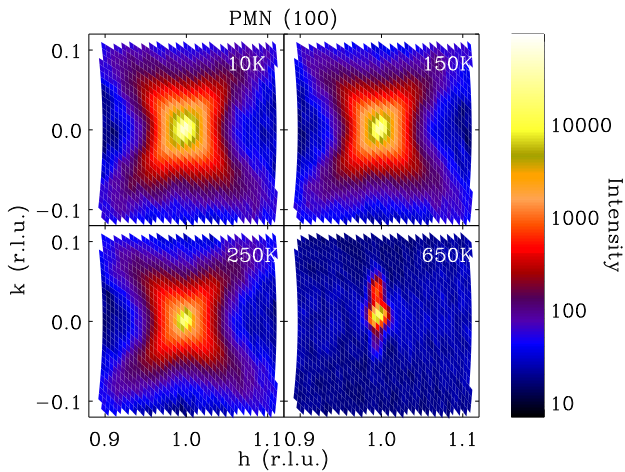}
\caption{Logarithmic plots of the neutron elastic diffuse  scattering intensity
around a (100) Bragg peak, at different temperatures.}
\label{fig:fig1}
\end{figure}

In Fig.~\ref{fig:fig1}, we show the neutron diffuse scattering 
around a (100) peak in the (hk0) plane measured 
at different temperatures. 
Compared with similar x-ray measurements, our neutron scattering measurements
have much higher energy resolution, and almost all phonon contributions can be
removed by integrating the scattering over an appropriate energy range, 
$0.075$~meV either side of the elastic peak position.
In fact, the energy spectra show that there is 
no significant scattering intensity outside this integral range, i.e.,
phonon contributions are indeed very small around the (100) Bragg peak.
The remaining low energy phonon contributions included within this
energy integral (phonons with energy transfer $|\hbar\omega| \le 0.075$~meV)
can also be  estimated, because phonon intensity increases with temperature, 
while the elastic diffuse scattering intensity decreases. Since no change in 
lineshape with temperature was observed, we believe that
phonon contributions to the elastic diffuse scattering intensity are
negligible.

\begin{figure}[ht]
\includegraphics[angle=90,width=\linewidth]{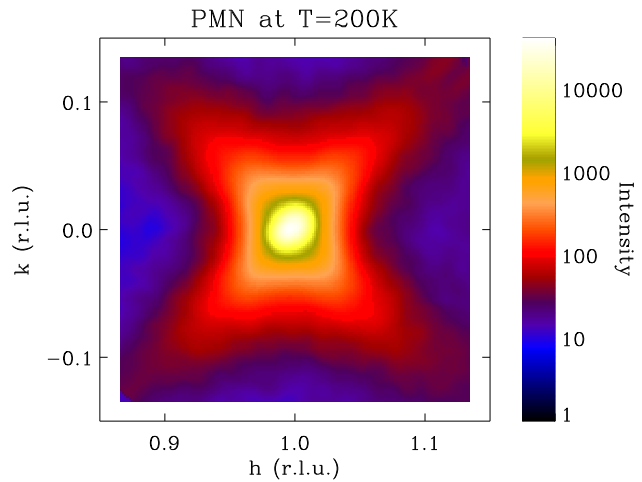}
\caption{A smoothed logarithmic plot of the neutron elastic diffuse 
scattering intensity at 200~K.}
\label{fig:fig2}
\end{figure}

Comparing the data sets plotted in Fig.~\ref{fig:fig1}, one can clearly 
see that diffuse scattering develops around the (100) peak 
with decreasing temperature. At $\sim 650$~K, the ``butterfly''-shaped diffuse 
scattering
intensity pattern observed at low temperatures is absent, and only a small 
trace of intensity transverse to the wave vector $Q=(1,0,0)$ remains. 
An enlarged plot of the diffuse scattering intensity measured at 200~K 
is given in Fig.~\ref{fig:fig2}. The diffuse scattering intensity has a 
``butterfly'' shape and extends from the Bragg peak along
(110) and (1$\bar{1}$0) directions. 
Previous x-ray diffuse scattering measurements~\cite{PMN_xraydiffuse2} show that
the polarizations of the PNR are along \{111\} directions in PMN, and the 
local symmetry is rhombohedral. This \{111\} type polarization 
gives the ``butterfly'' shaped diffuse intensity pattern in the (hk0) 
plane. The present elastic diffuse scattering results are 
in good agreement with the x-ray results.   
Recent neutron diffuse scattering measurements in the (hhk) 
scattering plane by S.-H.~Lee {\it et al.}~\cite{shl} provide additional 
evidence to support the \{111\} type polarization in PMN.

\begin{figure}[ht]
\includegraphics[width=\linewidth]{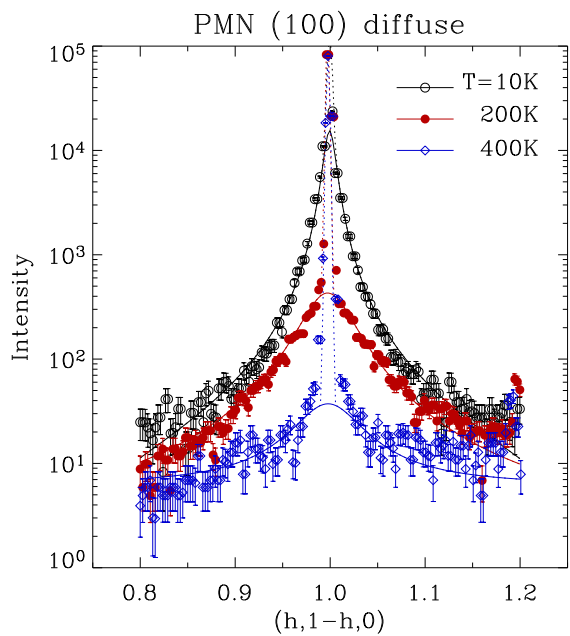}
\caption{Diffuse scattering intensities along the (110) direction 
near the (100) Bragg peak, measured at 10~K, 200~K, and 400~K. 
The intensity profiles  are fitted using a broad Lorentzian function (solid 
lines), a narrow Gaussian (dashed lines) that describes the resolution-broadened
Bragg peak, and a flat background.}
\label{fig:fig3}
\end{figure}

\begin{figure}[ht]
\includegraphics[width=\linewidth]{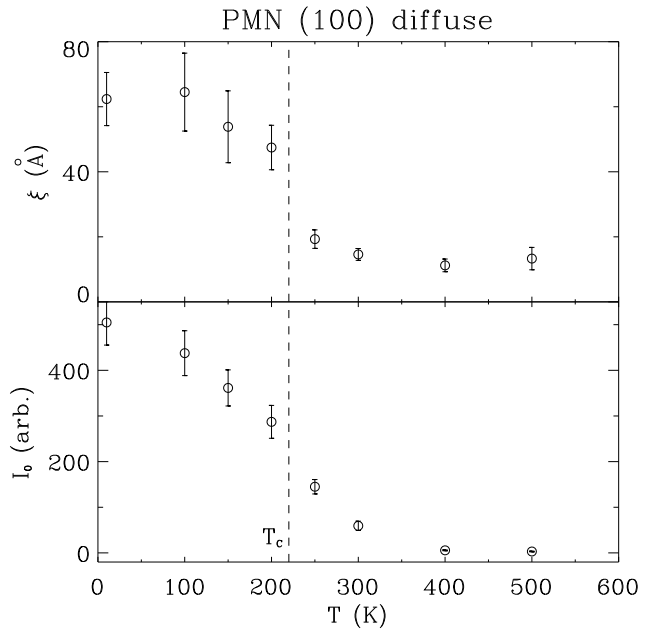}
\caption{Top frame: the correlation length $\xi$ as a function of T.  
Bottom frame:  integrated intensity $I_0$ as a function of T.}
\label{fig:fig4}
\end{figure}

Fig.~\ref{fig:fig3} shows cuts through the data 
along the (1$\bar{1}$0) direction at three selected temperatures. 
At 400~K, which is well above
$T_C$, the diffuse scattering intensity is already measurable. 
At higher temperatures, the diffuse scattering intensity 
becomes even weaker and harder to measure.  
We use a simple Lorentzian function to describe 
the spatial correlation of the atomic displacements contributing 
to the diffuse scattering: 
\begin{equation*}
I_{diff}=\frac{I_0\Gamma}{\pi(q^2+\Gamma^2)}, 
\end{equation*}
where $I_0$ is the integrated diffuse scattering intensity, 
$\Gamma=1/\xi$ is the inverse of 
the real space correlation length $\xi$, and $q$ is the length of the 
wave vector measured from the (100) Bragg position. Fig.~\ref{fig:fig3}
shows the results of fits to the Lorentzian, plus a Gaussian that describes the 
central Bragg peak, and a flat background. We have been able to obtain 
good fits using this model. The fitting 
parameters are the integrated intensity $I_0$ and the half width at half 
maximum $\Gamma$ of the Lorentzian, the intensity and width of the 
central Gaussian, and a flat background.
$I_0$ and $\xi=1/\Gamma$ are plotted in Fig.~\ref{fig:fig4}.

\begin{table*}[ht]
\caption{Integrated diffuse scattering intensity 
$I_0$, correlation length $\xi$, and the intensity at a selected wave vector
$Q=(1.05,0.05,0)$ vs temperature T.}
\begin{ruledtabular}
\begin{tabular}{lccccccccc}
T (K) & 10 &100&150&200&250&300&400&500&600\\
Integrated Intensity $I_0$ &504.9&437.6&361.4&287.1&145.0&59.3&
5.62&3.06&-\\
$\xi$ (\AA) &62.4&64.5&53.8&47.5&19.3&14.6&11.2&13.3&-\\
$I_{Q=(1.05,0.05,0)}$ &174.6&163.7&149.6&137.0&130.2&87.22&20.63&7.90&2.03\\
\end{tabular}
\end{ruledtabular}
\label{tab:1}
\end{table*}

In Table.~\ref{tab:1}, we list the integrated diffuse scattering intensity 
$I_0$, correlation length $\xi$, and the intensity at a selected wave vector
$Q=(1.05,0.05,0)$ vs temperature T.
The elastic diffuse scattering starts to be 
noticeable around $T \sim$400~K, much below $T_d\approx 620$~K. 
The ``correlation length'' $\xi$ is a direct measure of the length scale of
the static PNR. According to our results, the PNR are small when they first
appear at high temperatures, with average sizes around 
15~\AA~(see Fig.~\ref{fig:fig4}). Both the spatial correlation length of the 
atomic
displacements and the integrated intensity of the diffuse scattering increase
on cooling, even at temperatures below $T_C$ (Fig.~\ref{fig:fig4}).  At low
temperatures the length scale of the PNR reaches $\sim$65~\AA. 

\begin{figure}[ht]
\includegraphics[width=\linewidth]{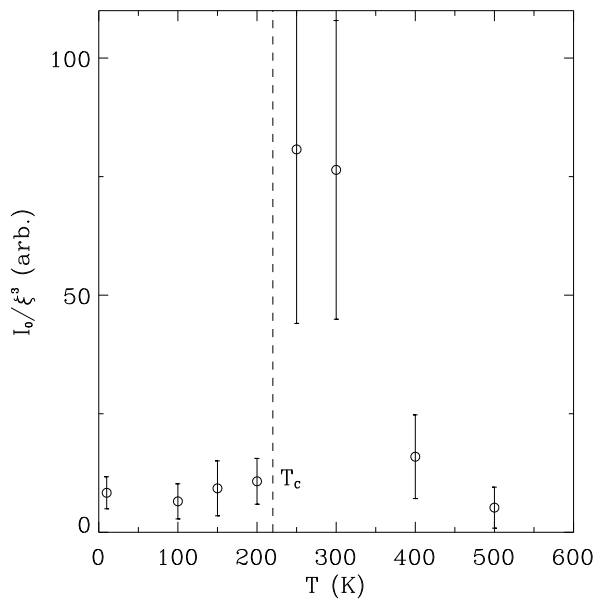}
\caption{Plot of $I_0/\xi^3$ vs T. Parameters were obtained from the fits to 
the diffuse data.}
\label{fig:fig5}
\end{figure}

From $T=300$~K to $100$~K, the volume of a single PNR ($V \propto \xi^3$) 
increases by a factor of $\sim$60, yet the integrated intensity only increases 
by a factor of $\sim$10. Writing $I_0$
as the product $N\xi^3|\bf{Q}\cdot\bf{\delta}|^2$, where $N$ is the 
total number of PNR
and $\delta$ is the average displacement of atoms within the PNR, we conclude
that $N|\bf{Q}\cdot\bf{\delta}|^2$ increases on cooling from high 
temperatures and then drops
dramatically at around $T_C$, remaining roughly constant below $T_C$. This is
illustrated in Fig.~\ref{fig:fig5} which shows $I_0/\xi^3$ as a function of
$T$. A likely scenario is that on cooling below $T_C$, $\delta$ remains
relatively constant whereas $N$ decreases as the smaller PNR merge
together.

Previous inelastic neutron scattering measurements on 
PMN~\cite{Waki1} have shown that the transverse optic (TO) mode phonon become 
overdamped  
near the zone center, starting at $T_d \approx 600$~K, which is roughly the 
temperature at which the PNR start to appear. We believe that the overdamping
of the soft TO mode phonon is directly associated with the formation of the 
PNR. In other words, the PNR may originate from the condensation of the soft 
TO mode. Reexamining the neutron diffuse data on PMN by Vakhrushev 
{\it et al.}~\cite{PMN_neutron3}, Hirota {\it et al.}~\cite{PMN_diffuse} 
have proposed a new interpretation of the atomic displacements derived from 
the data. The atomic displacements can be decomposed into
the sum of two terms with comparable magnitudes: 
$\delta_{soft}+\delta_{shift}$. $\delta_{soft}$ values satisfy the 
center-of-mass condition and are consistent with the values derived from
inelastic scattering intensities from the soft TO mode. The other term, 
$\delta_{shift}$, shows that the PNR are shifted along their polar 
directions relative to the surrounding lattices. This ``uniform phase 
shift'' has established convincing connections between the diffuse scattering 
from the PNR and the condensation of soft TO mode phonons. It is important to 
note that the magnitude of $\delta_{shift}$ is comparable to $\delta_{soft}$, 
which can be estimated to be around 1/10 of the lattice 
spacing. This large phase shift creates a huge energy barrier, preventing these 
polar regions from merging into the surrounding lattices. 

Recently, Gehring {\it et al.} have performed measurements 
that unambiguously established a true ferroelectric soft mode in 
PMN~\cite{PMN_softmode}. Further inelastic neutron scattering studies by 
Wakimoto {\it et al.}~\cite{Waki1} show that the soft mode extends to much 
lower temperatures below $T_C$, and that the phonon energy squared 
$(\hbar\omega_0)^2$, which
is inversely proportional to the dielectric constant $\epsilon$, 
increases linearly with decreasing temperature. 
These results represent a dynamical signature of a ferroelectric
phase below $T_C$. The phase below $T_C$ is now believed to be the recently 
discovered phase X~\cite{PZN_Xu,Xu1}, which exhibits an average cubic 
structure. However, unlike conventional ferroelectric phase transitions,
no macroscopic (rhombohedral) lattice distortion has been observed. The
decoupling of the lattice distortion from the ferroelectric 
polarization is quite intriguing, and we believe that the interaction between 
the PNR and the surrounding lattices is the key in solving this puzzle. 

Our current understanding of PNR in relaxor systems can be described as 
follows: 
At the Burns temperature $T_d$, the soft TO phonon mode becomes overdamped near 
the zone center, and starts to condense into PNR. These regions
form with local polarizations along \{111\} directions and are shifted
uniformly along their individual polarization direction.
The number and size of PNR in the system increases with cooling.
At the phase transition $T=T_C$, a large scale overall ``freezing'' of
the PNR occurs. Small PNR merge into larger ones and the total volume of 
PNR in the system keeps increasing. 
The related ferroelectric soft mode lifetime increases below $T_C$, 
and the overdamping near the zone center disappears. A macroscopic 
ferroelectric polar phase without lattice distortion is then 
established. Below $T_C$ the size of the PNR can grow slowly 
with further cooling. However, if the coupling between the PNR and the 
surrounding 
lattice is not sufficiently strong, as is the case in pure PMN, then the energy 
barrier 
created by the uniform phase shift would prevent the PNR from merging further  
and forming macroscopic lattice distortions.  The resulting phase will have a 
polar lattice of average cubic structure, but with embedded (rhombohedrally) 
polarized PNR. 

In addition to neutron and x-ray diffuse scattering measurements, recent Raman 
studies~\cite{Raman1} and specific heat measurements~\cite{PMN_heat} have also
provided useful information on PNR in PMN. We are planning to carry out 
further diffuse scattering measurements on a series of relaxor systems in the 
near future. 

In summary, we have shown high resolution neutron elastic  diffuse scattering 
data from PMN in the temperature range of 10~K to 650~K. The formation and 
development
of PNR at temperatures above, around, and below $T_C$ have been carefully 
studied. 
A merging of smaller PNR into larger ones occurs at the ferroelectric phase 
transition, while further growth of the PNR into macroscopic rhombohedral 
domains can not be achieved below $T_C$.

\section{Acknowledgments}

We would like to thank S.-H.~Lee, S.~B.~Vakhrushev,  D.~Viehland, and Z.-G.~Ye
for stimulating discussions. Financial support from the U.S. Department of 
Energy under contract No.~DE-AC02-98CH10886, and the National Science Foundation
under Agreement No.~DMR-0086210 is also gratefully acknowledged.

\end{document}